%% file: main.tex
\documentclass{article}

\usepackage{PRIMEarxiv}

\usepackage[utf8]{inputenc} 
\usepackage[T1]{fontenc}    
\usepackage{hyperref}       
\usepackage{url}            
\usepackage{booktabs}       
\usepackage{amsfonts}       
\usepackage{nicefrac}       
\usepackage{microtype}      
\usepackage{lipsum}
\usepackage{fancyhdr}       
\usepackage{graphicx}       
\graphicspath{{media/}}     

\usepackage{colortbl}

\usepackage[T1]{fontenc}
\usepackage{tcolorbox} 
\usepackage{geometry}  
\usepackage{listings}  
\usepackage{enumitem}  
\usepackage{xcolor}    
\usepackage{multirow}

\tcbset{
    chatbox/.style={
        colback=blue!5!white,
        colframe=blue!75!black,
        coltitle=black,
        fonttitle=\bfseries,
        fontupper=\small, 
        title={AI},
        sharp corners=southwest,
        boxrule=0.5mm,
        width=\columnwidth
    },
    human/.style={
        colback=gray!10!white,
        colframe=black!60!white,
        coltitle=black,
        fonttitle=\bfseries,
        fontupper=\small, 
        title={Human},
        sharp corners=southwest,
        boxrule=0.5mm,
        width=\columnwidth
    }
}

\title{KubeIntellect: A Modular LLM-Orchestrated Agent Framework for End-to-End Kubernetes Management
}

\author{
  Mohsen Seyedkazemi Ardebili, Andrea Bartolini \\
  Department of Electrical, Electronic, and Information Engineering \\
  University of Bologna \\
  Bologna, Italy\\
  \texttt{\{mohsen.seyedkazemi, a.bartolini\}@unibo.it} \\
}

\begin{document}
\maketitle

\begin{abstract}

Kubernetes has become the foundation of modern cloud-native infrastructure, yet its management remains complex and fragmented. Administrators must navigate a vast API surface, manage heterogeneous workloads, and coordinate tasks across disconnected tools—often requiring precise commands, YAML configuration, and contextual expertise.

This paper presents \textbf{KubeIntellect}, a Large Language Model (LLM)-powered system for intelligent, end-to-end Kubernetes control. Unlike existing tools that focus on observability or static automation, KubeIntellect supports natural language interaction across the full spectrum of Kubernetes API operations, including read, write, delete, exec, access control, lifecycle, and advanced verbs. The system uses modular agents aligned with functional domains (e.g., logs, metrics, RBAC), orchestrated by a supervisor that interprets user queries, maintains workflow memory, invokes reusable tools, or synthesizes new ones via a secure Code Generator Agent.

KubeIntellect integrates memory checkpoints, human-in-the-loop clarification, and dynamic task sequencing into a structured orchestration framework. \textbf{Evaluation results show a 93\% tool synthesis success rate and 100\% reliability across 200 natural language queries}, demonstrating the system's ability to operate efficiently under diverse workloads. An automated demo environment is provided on Azure, with additional support for local testing via \texttt{kind}. This work introduces a new class of interpretable, extensible, and LLM-driven systems for managing complex infrastructure.

\end{abstract}


\input{text/introduction}

\input{text/relatedwork}

\input{text/api}

\input{text/arch}

\input{text/implementation}
\input{text/evaluation}

\input{text/demo}
\section{Summary and Future Work}
\label{sec:conclusion}
The increasing complexity of Kubernetes cluster management presents significant challenges for administrators, demanding a high level of expertise and constant attention. Traditional tools often fall short due to their static nature and lack of adaptability. This paper introduces an innovative approach utilizing a Large Language Model (LLM) to revolutionize Kubernetes management. The proposed system integrates natural language processing with state-driven task orchestration to deliver intelligent automation and actionable insights. By leveraging a modular design, the system automates routine tasks such as log monitoring, error detection, and configuration analysis, while enabling real-time health monitoring of clusters. The LLM-powered framework serves as a supervisory layer, orchestrating task execution and optimizing workflows. Evaluations demonstrate the system's potential to reduce administrative burdens and improve decision-making, ultimately enhancing efficiency.

Future research will focus on evaluating the system's scalability and reliability in managing larger and more complex Kubernetes clusters. Comprehensive stress tests will be conducted to measure response times and resource utilization under high-demand scenarios. Additionally, the integration of advanced AI capabilities and extended support for DevOps tools will be explored to enhance the system's adaptability. Security enhancements for dynamic code execution and local deployment options for privacy-conscious environments, will be prioritized. Finally, further refinement of task orchestration strategies and real-time optimization techniques will be pursued to strengthen the system's performance in dynamic, production-grade environments.

\section{Acknowledgments}
Grammar checking and linguistic improvements are performed with the assistance of ChatGPT.

\bibliographystyle{unsrt}  
\bibliography{ref}

\input{text/appendix}
\end{document}

%% file: text/introduction.tex
\section{Introduction}
Kubernetes is the leading platform for orchestrating containerized applications in cloud-native environments. It enables scalable, declarative management of distributed systems across cloud, datacenter, and edge infrastructure. Features such as automatic scaling, self-healing, and service discovery have made it a cornerstone of modern platform engineering and DevOps workflows.
\subsection*{Challenges in Kubernetes Management}

Kubernetes offers powerful abstractions for managing containerized applications, but its complexity remains a barrier to efficient operation. Administrators must navigate a vast API surface, author intricate YAML configurations, and interact with low-level tools like \texttt{kubectl}, \texttt{k9s}, or Prometheus—all of which demand contextual expertise and precision. Small errors in command syntax, namespace targeting, or configuration logic can result in cascading failures or misconfigurations.

Day-to-day cluster management spans a wide range of tasks, including log inspection, resource scaling, configuration updates, access control enforcement, and in-cluster execution. These actions involve multiple resource types (e.g., Pods, Deployments, RBAC bindings) and diverse verbs—such as \texttt{get}, \texttt{patch}, \texttt{exec}, and \texttt{apply}—each with operational implications. While dashboards support observability, they often lack mutation and lifecycle controls, requiring users to bridge multiple tools manually.

This fragmented tooling ecosystem increases operational overhead and limits scalability. As organizations expand their Kubernetes footprints across clusters and teams, there is a growing demand for intelligent systems that abstract away low-level complexity, unify control and observability, and offer safe, interpretable interfaces to the full Kubernetes API surface.

\subsection*{The Role of LLMs in Orchestration and Control}

Recent advances in large language models (LLMs) have opened new possibilities for interacting with complex systems through natural language. LLMs can interpret ambiguous human queries, generate structured outputs, and reason over task sequences—making them attractive candidates for intelligent system interfaces. In the context of Kubernetes, LLMs offer the potential to abstract the intricacies of API verbs, resource schemas, and control workflows by translating natural language instructions into actionable plans.

Unlike traditional automation scripts or static dashboards, LLMs can dynamically adapt to user intent, reformulate queries, and integrate contextual hints during execution. When embedded within a task orchestration framework, LLMs can serve as a control hub—selecting appropriate tools, invoking agents, or even generating new scripts on demand. This approach enables a more flexible and interactive model of Kubernetes management, where users describe \textit{what} they want to achieve, and the system determines \textit{how} to do it safely and effectively.

However, deploying LLMs in operational infrastructure requires more than prompt-response generation. It demands structured reasoning, memory, access control, and safe execution environments. LLMs must be grounded in a secure, modular architecture that supports validation, fallback, policy enforcement, and integration with real-world APIs. This motivates the design of orchestration frameworks that combine natural language understanding with formal task planning, agent-based execution, and robust infrastructure controls.

\subsection*{Conceptual Definitions and Scope}

To understand the scope of this work, several core concepts must be defined. At the heart of Kubernetes management are its \textbf{API verbs}, which define permissible operations on cluster resources. These include \texttt{get}, \texttt{list}, and \texttt{watch} for reading state; \texttt{create}, \texttt{update}, \texttt{patch}, and \texttt{delete} for modifying resources; \texttt{exec}, \texttt{attach}, and \texttt{proxy} for executing commands inside containers; and additional verbs such as \texttt{scale}, \texttt{bind}, and \texttt{approve} for lifecycle and policy control. Together, these verbs define the full \textit{control surface} of Kubernetes.

In this context, an \textbf{agent} refers to a modular subsystem responsible for a specific category of Kubernetes operations—such as log analysis, metrics gathering, configuration inspection, access auditing, or workload scaling. Each agent encapsulates reusable tools aligned with its domain and interacts with the Kubernetes API to execute structured tasks. Tools can be either predefined (hand-coded and tested) or synthesized dynamically from natural language requests.

\textbf{Code generation} refers to the system’s ability to create new tools at runtime using a prompt-engineered LLM. When a user request does not match any known tool, the system triggers a pipeline that generates, validates, and registers a Python script implementing the requested functionality. This script is executed in a secure sandbox and must pass functional and structural tests before becoming part of the available agent toolkit.

\textbf{Human-in-the-loop (HITL)} interaction is used in cases where the LLM identifies ambiguity or uncertainty in a user query or tool output. The system prompts the user for clarification or approval before proceeding, ensuring safe and interpretable decision-making.

\textbf{Task orchestration} is the process of converting a user’s intent into an executable, multi-step plan. This is achieved through a finite-state workflow engine that routes tasks across agents, manages execution memory, supports conditional branching, and integrates checkpoints for resumability.

Together, these elements support what we refer to as \textbf{end-to-end orchestration across the Kubernetes control surface}—a model in which natural language queries can trigger complex workflows that span from observability to mutation, access policy checks, lifecycle actions, and dynamic tool generation.

\subsection*{Research Gap}

Despite the growing ecosystem of Kubernetes tools—including CLI wrappers, web-based dashboards, and observability suites—most solutions remain limited in scope and flexibility. They are primarily designed for visualizing state and assisting with routine diagnostics, offering only partial support for the broader set of Kubernetes operations.

First, the majority of these tools are constrained to \textbf{read-only observability}, focusing on log inspection, metric dashboards, and resource querying. They provide limited capabilities for \textbf{control-plane actions} such as modifying configurations, scaling workloads, evicting pods, or applying new manifests—operations that are central to day-to-day cluster management.

Second, existing platforms generally lack support for \textbf{coordinated multi-step workflows}. Many administrative tasks—such as retrieving metrics, evaluating them against custom rules, and then triggering remediation—require chaining multiple Kubernetes API calls. Without a structured orchestration layer, users must manually bridge these steps using disconnected tools or write custom scripts.

Third, current systems offer \textbf{no mechanism for dynamic tool synthesis}. When a desired capability is not available—e.g., querying across namespaces with fine-grained filters—users are required to write and maintain ad hoc scripts or definitions, increasing the potential for human error and limiting agility.

Finally, we observe an absence of intelligent orchestration frameworks that incorporate \textbf{language-based intent interpretation}, contextual reasoning, execution memory, and human-in-the-loop (HITL) interaction. Existing systems do not support shared state across workflows, checkpointed execution, or clarification mechanisms that respond dynamically to ambiguous or evolving user queries. Moreover, natural language processing is rarely used to mediate user intent or drive execution paths, leaving a gap in how administrators interact with complex infrastructure.

This paper addresses these limitations by introducing a novel system that integrates natural language interaction, modular agents, structured task orchestration, and secure code generation. The proposed architecture enables comprehensive, context-aware, and extensible management of Kubernetes environments—supporting both observability and control tasks within a unified interface.

\subsection*{Main Contributions of This Work}

This paper introduces \textbf{KubeIntellect}, a novel LLM-powered framework for Kubernetes management that combines modular design, dynamic tool generation, and explainable orchestration. The system supports natural language interaction and enables end-to-end automation across all major Kubernetes control domains.

The key contributions of this work are as follows:

\begin{itemize}
    \item \textbf{A modular, multi-agent architecture} that abstracts Kubernetes operations into specialized agents aligned with functional domains such as logging, configuration, metrics, security, and lifecycle management.
    
    \item \textbf{Comprehensive support for the full Kubernetes API surface}, encompassing all seven categories of API verbs: read, write/modify, delete, exec/proxy, permission/auth, scale/lifecycle, and custom/advanced operations.
    
    \item \textbf{A dynamic Code Generator Agent} that synthesizes new tools from natural language descriptions, validates them 
    , and registers them into the agent ecosystem with metadata and audit support.
    
    \item \textbf{A LangGraph-based orchestration engine} that enables structured, explainable workflows with support for conditional execution, persistent memory, human-in-the-loop clarification, and task resumption via PostgreSQL-backed checkpoints.
    
    \item \textbf{End-to-end automation of operational tasks}, ranging from querying resource state to modifying workloads, enforcing access policies, and executing contextual remediation—entirely through a natural language interface.
    
    \item \textbf{A reproducible, cloud-deployable testing environment} for KubeIntellect using Azure Kubernetes Service (AKS), along with early support for local testing via \texttt{kind}, enabling wide accessibility and rapid onboarding.
\end{itemize}

\subsection*{Paper Organization}

The remainder of this paper is structured as follows:

Section~\ref{sec:state_of_the_art} reviews related work in Kubernetes automation and intelligent system interfaces.  
Section~\ref{sec:api_scope} introduces the Kubernetes API capability spectrum and categorizes the operational domains addressed by KubeIntellect.  
Section~\ref{sec:architecture} presents the system architecture, detailing its layered design, agent modules, orchestration logic, and memory model.  
Section~\ref{sec:implementation} describes the implementation, including the use of LangGraph, structured tools, and sandboxed code execution.  
Section~\ref{sec:evaluation} evaluates the system through latency analysis, workload behavior, and resource usage across multiple test environments.  
Section~\ref{sec:availability} outlines code availability, cloud deployment infrastructure, and ongoing support for local testing.  
Finally, Section~\ref{sec:conclusion} concludes the paper and outlines directions for future work.

%% file: text/relatedwork.tex
\section{State of the Art}
\label{sec:state_of_the_art}

Recent advancements in the intersection of Large Language Models (LLMs) and infrastructure automation have given rise to various systems for configuration generation, failure remediation, and agent-based orchestration. These works demonstrate the growing applicability of LLMs in system administration but differ in scope, depth, and operational focus.

\subsection{LLM-based Configuration Tools}

LADs~\cite{khan2025lads} explores the use of LLMs to generate configuration files for distributed systems like Dask, Redis, and Ray. It applies few-shot learning, retrieval-augmented generation (RAG), and prompt chaining to produce valid deployment configurations. LADs contributes to DevOps by automating initial setup procedures but does not address runtime infrastructure management or dynamic orchestration of live systems.

\subsection{Role-Aware Multi-Agent Architectures}

AgentFM~\cite{zhang2025agentfm} proposes a role-aware multi-agent system for failure detection and remediation in distributed databases. By assigning agents to system, data, and task roles, it improves anomaly management in platforms like TiDB and IoTDB. The framework highlights the benefits of specialized agents and LLM reasoning in constrained domains. However, its design is specific to database infrastructure and does not generalize to broader orchestration or real-time control across cloud-native environments.

\subsection{Autonomous Operations and Agent Benchmarking}

AIOpsLab~\cite{shetty2024building, chen2025aiopslab} introduces a benchmarking platform to evaluate the performance of LLM agents in simulated cloud incident scenarios. Using tools such as Prometheus and ChaosMesh, AIOpsLab provides structured environments for testing recovery strategies. These systems focus on observability and fault injection, offering valuable datasets and metrics. Nevertheless, they stop short of enabling real-time, user-driven automation of infrastructure tasks.

\subsection{SLM Agents for Procedural Safety}

Authors of ~\cite{fukuda2025small} investigate the use of Small Language Models (SLMs) for safe execution of ICT operations, emphasizing procedural reliability in resource-constrained environments. Their framework employs dynamic prompt engineering, exemplar selection, and nested reasoning to prevent shortcut errors. While this method offers lightweight and privacy-preserving capabilities, it primarily targets step-by-step procedure following rather than complex, modular orchestration workflows.

\subsection{General-Purpose Agent Frameworks and IaC Systems}

AutoAgent~\cite{tang2025autoagent} enables non-technical users to create LLM-powered agents through natural language alone. It emphasizes accessibility, offering modular workflow generation and API integration for a broad range of tasks. Similarly, IaCGen~\cite{zhang2025deployability} focuses on generating deployable Infrastructure-as-Code (IaC) templates using LLMs with deployment validation feedback loops. These systems are designed for general agent usability or static template generation, not for executing or managing running infrastructure in real-time.

\subsection{Limitations in Prior Work}

Across these diverse approaches, several key limitations emerge:
\begin{itemize}
    \item \textbf{Narrow Scope}: Many systems focus on configuration generation or fault handling in specific domains (e.g., databases, static IaC).
    \item \textbf{Lack of Runtime Orchestration}: Few frameworks support dynamic execution or control across the full surface of infrastructure APIs.
    \item \textbf{Limited Extensibility}: Most works do not support dynamic tool generation or pluggable agent architectures.
    \item \textbf{Absence of Memory and HITL}: Persistent workflow memory, checkpointing, and human-in-the-loop (HITL) clarification are often missing.
    \item \textbf{Minimal Security Integration}: Few systems incorporate secure execution environments, RBAC enforcement, or auditability.
\end{itemize}

\subsection{KubeIntellect: Bridging the Gap}

To address these gaps, we propose \textbf{KubeIntellect}, a domain-specialized LLM-powered orchestration system for Kubernetes management. Unlike prior works that focus on individual aspects such as configuration or monitoring, KubeIntellect introduces:
\begin{itemize}
    \item A modular, multi-agent architecture aligned with Kubernetes API domains (e.g., logs, metrics, RBAC, scaling, configs).
    \item LangGraph-based orchestration with persistent memory, checkpointing, and workflow resumption.
    \item A dynamic code and tool generation engine with sandboxed execution and audit logging.
    \item Built-in human-in-the-loop (HITL) decision points for safe execution of ambiguous or risky operations.
    \item Compatibility with both cloud-hosted and local LLMs via a unified abstraction layer.
\end{itemize}

Through this architecture, KubeIntellect transforms natural language queries into interpretable, secure, and executable workflows that support the full Kubernetes lifecycle—from diagnostics and configuration to runtime control and remediation. This positions KubeIntellect as a significant step forward in explainable and adaptive infrastructure automation.

%% file: text/api.tex
\section{Operational Scope of KubeIntellect}
\label{sec:api_scope}
Effective Kubernetes management spans far beyond traditional observability. The Kubernetes API surface includes operations for resource querying, workload creation, configuration updates, in-cluster execution, access control validation, and lifecycle control. KubeIntellect is designed to support this full range of operational categories through both pre-defined agents and dynamically synthesized tools.

Table~\ref{tab:k8s_api_operations} summarizes the primary categories of Kubernetes API actions supported by KubeIntellect. These span both read- and write-oriented verbs and reflect real-world administrative operations used in cluster management.

\begin{table}[]
\centering
\caption{Categories of Kubernetes API Operations Supported by KubeIntellect}

\label{tab:k8s_api_operations}
\resizebox{\textwidth}{!}{%
\begin{tabular}{lll}
\hline
\multicolumn{1}{c}{\textbf{Category}} & \multicolumn{1}{c}{\textbf{API Verbs}}          & \multicolumn{1}{c}{\textbf{Example Operations}} \\ \hline
\textbf{Read}                         & \texttt{get}, \texttt{list}, \texttt{watch}                                & List pods, get pod logs, watch events           \\
\textbf{Write / Modify}               & \texttt{create}, \texttt{update}, \texttt{patch}, \texttt{replace}                  & Create a deployment, patch configmap            \\
\textbf{Delete}                       & \texttt{delete}, \texttt{deletecollection}                        & Delete a job, delete all pods in a namespace    \\
\textbf{Execute / Proxy}              & \texttt{exec}, \texttt{port-forward}, \texttt{attach}, \texttt{proxy}, \texttt{log}          & Run in-container commands, port forwarding, logs        
\\
\textbf{Permission \& Auth}           & \texttt{impersonate}, \texttt{authorize}, 
& Validate RBAC, perform access reviews               \\
\textbf{Scale / Lifecycle}            & \texttt{scale}, \texttt{rollout}, \texttt{restart}, \texttt{evict}, \texttt{cordon}          & Scale workloads, restart pods, drain nodes \\
\textbf{Custom / Advanced}            & \texttt{apply}, \texttt{finalize}, \texttt{bind}, \texttt{approve}, \texttt{renew}           & Apply manifests, bind pods, approve certificates         
\\ \hline
\end{tabular}
}
\end{table}

By supporting all of these categories, KubeIntellect offers a comprehensive operational interface—not limited to read-only observability, but enabling full control over the Kubernetes environment. This includes advanced actions such as automated scaling, configuration patching, node draining, and interactive command execution.

For detailed breakdowns of how monitoring-specific data types (e.g., logs, metrics, configurations, and security events) are processed within this framework, see Appendix~\ref{appendix:monitoring_data}.


%% file: text/arch.tex
\section{System Architecture}
\label{sec:architecture}

\subsection{Overview and Design Principles}
\begin{figure}[htbp]
\centering
\includegraphics[width=0.8\linewidth]{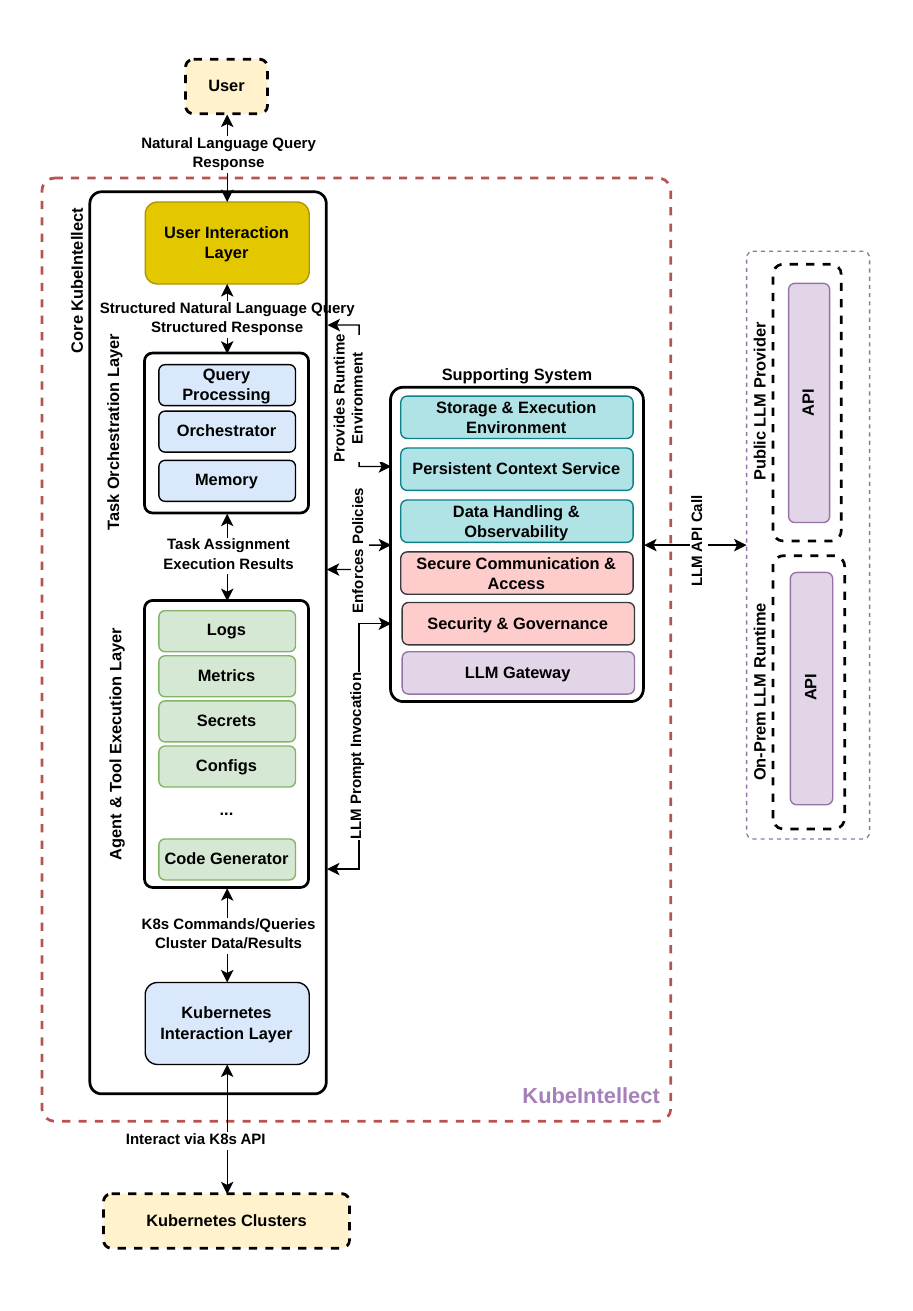}
\caption{High-level architecture of the KubeIntellect system. The core system (left column) is structured into four primary layers: (1) the User Interaction Layer, which create GUI for user natural language queries; (2) the Task Orchestration Layer, powered by an LLM for reasoning, memory, and coordination; (3) the Agent and Tool Execution Layer, where modular agents handle domain-specific Kubernetes tasks; and (4) the Kubernetes Interaction Layer, which executes API calls to the cluster. The supporting system (right column) includes persistent memory, execution infrastructure, observability, governance, and model abstraction via the LLM Gateway. External LLM runtimes (public or private) are accessed through a unified API interface. This architecture enables secure, explainable, and modular automation of Kubernetes management workflows.}
\label{fig:systemarchitecture}
\end{figure}

KubeIntellect is an LLM-powered, multi-agent system designed to simplify Kubernetes cluster management through natural language interaction and intelligent orchestration. The architecture follows a modular and layered design philosophy, where each component is dedicated to a specific responsibility—from interpreting user queries to executing Kubernetes operations securely and reliably. Central to the system is a Large Language Model (LLM) that functions as a reasoning engine, coordinating task execution across specialized agents and dynamically adapting workflows through a LangGraph-based orchestration mechanism. The design prioritizes extensibility, allowing new agents, tools, or models to be integrated with minimal reconfiguration. Security and compliance are embedded throughout the stack, supported by infrastructure components for policy enforcement, RBAC, sandboxed code execution, and secure communication. The system also includes a persistent context service for memory and checkpointing, enabling continuity across sessions. A comprehensive architectural overview of KubeIntellect is illustrated in Figure~\ref{fig:systemarchitecture}.

\subsection{Core KubeIntellect System}
\subsubsection{User Interaction Layer: }
The User Interaction Layer serves as the primary interface between end-users and the KubeIntellect system. It abstracts the complexities of Kubernetes operations by allowing users to issue queries in natural language, without requiring familiarity with kubectl commands, resource definitions, or API structures. This layer captures user input through a chat-like interface, which may be integrated into a web frontend or terminal-based assistant. It ensures a smooth interaction loop by forwarding the input to the 
query processing module (at Task Orchestration Layer) and rendering the final structured response in a human-readable format. Output may include structured tables, status summaries, or error reports, depending on the task type. By decoupling users from the intricacies of Kubernetes internals, this layer plays a critical role in simplifying cluster management and enhancing operator productivity.

\subsubsection{Query Processing Module: }
The Query Processing Module serves as the semantic and policy-aware entry point of KubeIntellect. Its primary role is to perform early-stage validation, filtering, and coarse-grained interpretation of user queries before they are passed to the orchestration engine. Leveraging a Large Language Model (LLM), this module identifies unsupported requests—such as greetings, irrelevant or out-of-domain questions, malformed input, or encoded gibberish—and rejects or reformulates them as necessary. It also applies role-based constraints, denying privileged actions (e.g., creating pods or secrets) for users with read-only access profiles. Ambiguous queries may trigger clarification prompts, while semantically incomplete requests are normalized to ensure consistent downstream processing.

The output of this module is a structured, validated query object that captures the user’s intent, scope (e.g., namespaces, pods), and any contextual hints. However, it does not constitute a complete execution plan. This intermediate representation is then forwarded to the Task Orchestration Module.
By isolating interpretation and compliance checking in a dedicated module, KubeIntellect ensures robust query handling, enforces access policies, and simplifies the downstream orchestration logic.

\subsubsection{Task Orchestration Module: }
The Task Orchestration Module is the central decision-making and coordination unit within KubeIntellect, responsible for converting validated queries into executable workflows. It receives structured query objects from the Query Processing Module and transforms them into task plans, leveraging a finite-state machine (FSM). This FSM governs the dynamic routing of tasks through specialized agents such as Logs, Metrics, or Code Generator, based on the query type, system state, and intermediate outcomes.

The orchestrator maintains workflow context across steps using a persistent memory mechanism, enabling iterative refinement, human-in-the-loop checkpoints, and conditional branching. It dynamically evaluates agent responses, adjusts execution paths, and ensures that multi-step tasks—such as diagnosing cluster anomalies or retrieving filtered performance metrics—are completed correctly and transparently. 
By encapsulating control flow, agent coordination, and execution sequencing within a modular framework, the Task Orchestration Module ensures that KubeIntellect remains extensible, explainable, and responsive to both user intent and runtime data.

\begin{figure}[ht]
\centering
\includegraphics[width=\textwidth]{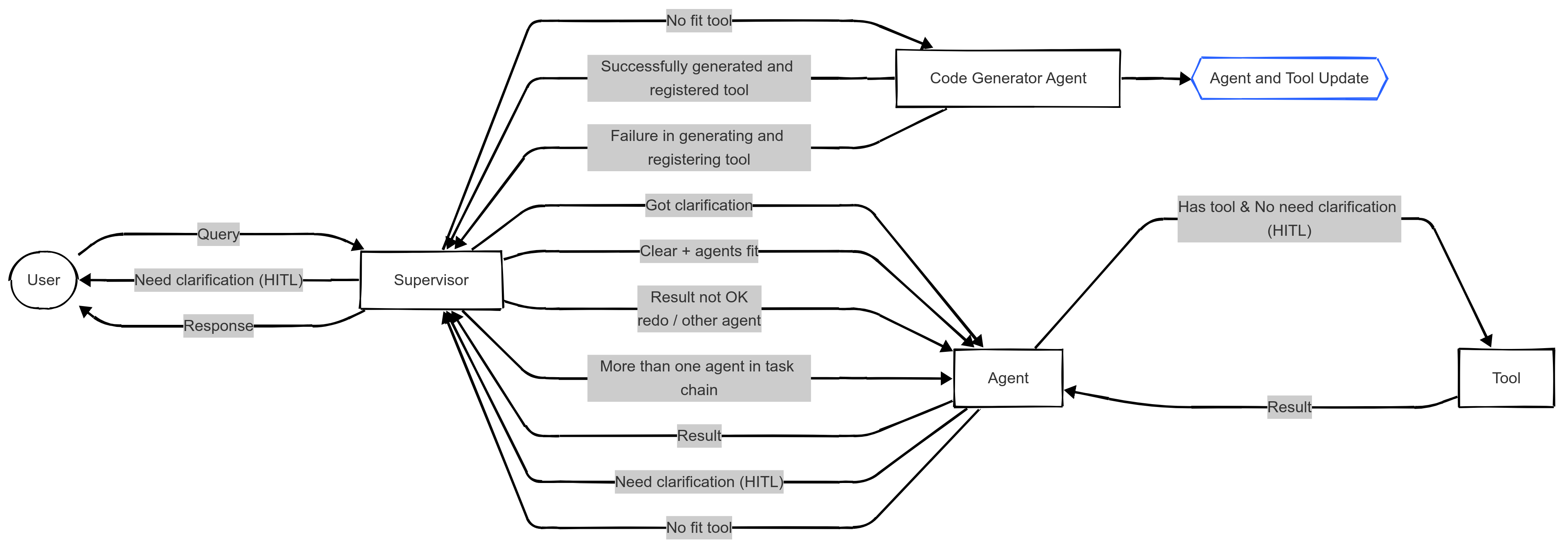}
\caption{%
Supervisor-driven decision flow in KubeIntellect. The diagram shows how user queries are parsed by the orchestrator, routed through available agents, and either resolved using existing tools or escalated to the Code Generator Agent. Human-in-the-loop clarification and retry logic are incorporated to ensure interpretability, correctness, and fallback resilience. This architecture supports both reusable automation and adaptive tool creation.%
}
\label{fig:supervisor-decision-flow}
\end{figure}

Figure~\ref{fig:supervisor-decision-flow} illustrates the Supervisor’s reasoning workflow, including agent selection, tool matching, human-in-the-loop (HITL) prompts, and fallback to dynamic code generation. This diagram provides a visual summary of the orchestration logic described above.

\subsubsection{Memory Module: }
\paragraph{Memory Module}
The Memory Module within KubeIntellect plays a pivotal role in enabling persistent, context-aware task execution across multi-step workflows. It is tightly integrated with the Task Orchestration Module through LangGraph’s memory and checkpointing primitives, which allow the orchestrator to maintain state throughout the lifecycle of a user query. 

To support this functionality, KubeIntellect employs a dual-memory strategy: ephemeral in-memory storage for short-lived orchestration state and a PostgreSQL-backed checkpoint store for durable persistence. The ephemeral memory retains intermediate results and workflow metadata within the context of a running state graph, enabling conditional branching, dependency resolution, and task rollback. Meanwhile, the checkpoint database captures execution snapshots at predefined decision points, enabling resumption of workflows after interruption, user approval checkpoints, and audit logging of LLM decisions and agent actions.

This hybrid approach ensures that agents have access to both current task context and historical decisions, enabling them to avoid redundant actions and produce consistent, explainable outcomes. It also supports human-in-the-loop interactions and multistep reasoning by preserving memory across sessions. By using the memory in the orchestration logic, KubeIntellect achieves robust continuity, task traceability, and enhanced decision-making in Kubernetes environments.

\subsubsection{Agent \& Tool Execution Layer: }
The Agent \& Tool Execution Layer comprises a collection of specialized agents, each tailored to a distinct category of Kubernetes operations such as log analysis, configuration inspection, metrics monitoring, security auditing, and dynamic tool generation. 
These agents encapsulate domain-specific logic and serve as autonomous execution units that receive task directives from the orchestrator and translate them into actionable Kubernetes API calls. Each agent is equipped with a suite of modular tools that abstract direct interactions with the Kubernetes API, supporting operations like listing pods, retrieving resource usage statistics, analyzing logs, or validating deployment configurations. The modular design of these tools (within each agent) ensures reusability, ease of extension, and composability across workflows. 
This layer enables targeted execution of complex tasks while maintaining separation of concerns, and serves as the operational backbone of the system by using the Kubernetes interaction layer to communicate with the control plane.

\input{text/code_gene}

\subsubsection{Kubernetes Interaction Layer}
The Kubernetes Interaction Layer serves as the secure interface between the KubeIntellect system and the underlying Kubernetes clusters. It is responsible for dispatching Kubernetes API calls issued by the agents, covering both read and write operations. These include retrieving pod logs, deployment configurations, and resource metrics, as well as creating or modifying resources such as pods, deployments, secrets, and jobs. Communication with the Kubernetes API server is established through a secure channel, typically an SSH tunnel, which ensures encrypted, authenticated, and isolated access—particularly important in multi-tenant or air-gapped environments.

This layer handles request routing, error propagation, and response normalization, providing a consistent and abstracted interface for the agent layer above. It enforces access control policies using Kubernetes-native mechanisms like RBAC, ensuring that all operations respect the permissions defined for each user or agent role. By encapsulating the intricacies of Kubernetes API interactions, this layer ensures that the execution logic remains agnostic to cluster topology, version, or authentication mechanism, while supporting both observability and control use cases within the cluster.

\subsection{Supporting Infrastructure Systems}

\subsubsection{LLM Integration Pathway}

\paragraph{LLM Gateway: }
The LLM Gateway acts as an intermediary layer between the core orchestration logic and underlying language model providers. Its primary role is to abstract provider-specific details, enabling seamless switching between different backends such as OpenAI, Azure OpenAI, or self-hosted LLM runtimes like LLaMA and Mistral. This abstraction simplifies integration by exposing a uniform interface for issuing prompts and handling completions, regardless of the backend.

Beyond abstraction, the gateway implements robustness features such as retry logic, error handling, and rate-limiting compliance to ensure reliability in the face of transient API failures or quota limits. It may also incorporate response caching to reduce latency for repeated prompts and enhance system efficiency. By decoupling the orchestrator from the specifics of model deployment and API semantics, the LLM Gateway promotes modularity, scalability, and flexibility in deploying KubeIntellect across diverse infrastructure environments.

\paragraph{LLM Providers (Public and Self-Hosted): }
KubeIntellect is designed to support both public LLM services and self-hosted model deployments, offering flexible deployment options to meet varying operational, regulatory, and performance requirements. Public providers such as OpenAI, Azure OpenAI, and Anthropic deliver managed APIs for state-of-the-art language models with minimal infrastructure overhead, making them ideal for scenarios prioritizing rapid integration and high model quality.

Conversely, KubeIntellect also supports self-hosted or privately deployed LLMs, including open-source models like LLaMA, Mistral, or GPT-J. These models can be hosted in private cloud environments or on-premises using inference engines such as Triton or vLLM or Hugging Face’s Text Generation Inference (TGI). This deployment mode offers greater control over data privacy, cost, latency, and model customization—particularly useful in secure, air-gapped, or compliance-sensitive environments.

Through the LLM Gateway, both public and private models are integrated using a unified abstraction layer that normalizes prompt-response formats and enforces shared policies. This architecture enables organizations to seamlessly switch between or combine providers without modifying the orchestration logic, ensuring consistent behavior across diverse LLM backends.

\subsubsection{Secure Communication and Access}
KubeIntellect enforces secure communication and access control through a combination of encrypted network tunnels, identity-aware access policies, and Kubernetes-native security mechanisms. Access to the Kubernetes API server is established via an SSH tunnel, which provides an encrypted communication channel between the KubeIntellect runtime and remote or air-gapped clusters. This approach isolates control plane access from public exposure, reducing the attack surface and enabling secure operations across network boundaries. In parallel, the system leverages Kubernetes Role-Based Access Control (RBAC) to restrict agent permissions according to the principle of least privilege. Each agent operates under a service account with predefined roles, ensuring that only authorized actions can be performed within the cluster. Identity enforcement mechanisms are applied consistently across the orchestrator, agents, and tools, enabling granular policy definition and auditability. This layered security model ensures confidentiality, integrity, and accountability in all interactions with the Kubernetes environment.

\subsubsection{Security and Governance}
The Security and Governance module in KubeIntellect enforces compliance, accountability, and operational integrity throughout the system. A key component of this module is the auditing subsystem, which records all significant interactions, including user queries, agent actions, API calls, and LLM outputs. These audit logs support forensic analysis, traceability, and adherence to governance standards in multi-tenant or regulated environments. In addition, the system includes a secure sandboxing mechanism for executing dynamically generated code—particularly from the Code Generator agent. This sandbox enforces strict isolation boundaries, limiting filesystem access, network communication, and execution time to mitigate risks associated with arbitrary code execution. Access monitoring mechanisms further enhance the system's security posture by detecting and alerting on anomalous behaviors, unauthorized access attempts, or policy violations. Together, these capabilities ensure that intelligent automation remains secure, controlled, and auditable in production-grade Kubernetes environments.

\subsubsection{Persistent Context Service}
The Persistent Context Service in KubeIntellect provides long-term memory and state management across multi-step workflows and user sessions. It leverages PostgreSQL-based checkpointing, integrated with LangGraph, to capture intermediate states, execution metadata, and contextual variables at various stages of a task. This enables the system to support features such as human-in-the-loop approval, workflow resumption, and multi-turn interactions without loss of state. Checkpoints store not only orchestration-level decisions but also agent outputs, tool responses, and derived insights that may influence downstream steps. The persistent context is accessible across agents, allowing them to share information, avoid redundant computations, and adapt behavior based on historical interactions. This shared memory architecture enhances reasoning continuity, reduces latency, and forms the basis for implementing traceable, explainable, and context-aware automation in Kubernetes environments.

\subsubsection{Data Handling and Observability}
Data handling and observability are critical to ensuring the reliability, debuggability, and performance of the KubeIntellect system. This subsystem provides data pipelines that collect, process, and deliver logs, and metrics 
from the Kubernetes environment to the monitoring and logging system. These pipelines enable near-real-time access to operational data, facilitating use cases such as anomaly detection, resource optimization, and compliance verification. In addition, KubeIntellect incorporates system health monitoring components that continuously assess the status of internal services—such as the LLM gateway, orchestrator, and checkpointing engine—and external dependencies like the SSH tunnel and Kubernetes API server. Health probes, heartbeat signals, and telemetry are used to trigger automated recovery mechanisms or alert system operators. Together, these capabilities provide deep visibility into both cluster-level dynamics and internal system behavior, enabling robust and self-aware orchestration.

\subsubsection{Storage and Execution Environment}
The Storage and Execution Environment provides the runtime foundation for executing agents, tools, and dynamically generated code within KubeIntellect. It ensures that all components operate within secure, reproducible, and isolated environments while enabling seamless state sharing and memory persistence.

Each agent and module runs in a containerized environment managed by Kubernetes, ensuring dependency consistency, process isolation, and portability across deployment platforms. Shared volumes and persistent volume claims (PVCs) are mounted to facilitate data exchange between agents and long-term storage of artifacts, logs, and intermediate results. These shared paths also support multi-agent workflows where data continuity between steps is critical.

Dynamic code execution—especially scripts generated by the Code Generator Agent—is handled within a secure REPL sandbox. This sandbox is governed by strict execution constraints, including CPU/memory limits, restricted I/O, and runtime timeouts. These safeguards ensure that user-generated or LLM-generated code executes safely without compromising the integrity of the host system.

The environment further integrates with KubeIntellect’s persistent context system via a PostgreSQL-backed checkpointing mechanism. This enables agents to access and update shared memory state, including workflow metadata, intermediate outputs, and orchestration decisions. It ensures that even across distributed or delayed workflows, execution state remains consistent and recoverable.

Dependency management is enforced via Poetry or pip, ensuring deterministic builds and controlled upgrades of language runtimes. Together, these capabilities enable a robust, secure, and extensible runtime system that supports both predefined agents and adaptive, user-driven toolchains in a unified manner.

\subsection{End-to-End Workflow Example}
To illustrate the interaction between components in KubeIntellect, consider the user query: \textit{“List all pods and identify those with errors in each namespace.”} The query is first submitted through the User Interaction Layer, which captures the input and forwards it to the Query Processing Module. This module uses an LLM to interpret the intent and generate a structured task plan that specifies a need to retrieve pod listings and analyze logs across all namespaces. The structured plan is handed over to the Task Orchestration Layer, which uses a LangGraph-based finite-state machine to sequence the tasks. The orchestrator first activates the Configs Agent, which invokes the Kubernetes Interaction Layer to list all pods in each namespace. The results are stored in the Persistent Context Service and passed back to the orchestrator, which dynamically routes the next step to the Logs Agent. This agent fetches logs for each pod and scans them for error patterns, again via the Kubernetes Interaction Layer. The orchestrator then aggregates the findings, stores the final output, and returns it to the User Interaction Layer. The user receives a structured response showing all pods along with a diagnosis of those exhibiting error conditions. This example highlights KubeIntellect’s ability to coordinate multi-agent workflows, access real-time cluster data securely, and produce explainable, actionable insights using natural language.

\subsection{Modularity and Extensibility}
KubeIntellect is architected with modularity and extensibility as core design principles, enabling seamless adaptation to evolving operational requirements and system contexts. The architecture separates concerns across well-defined layers—such as user interaction, orchestration, execution, and infrastructure—ensuring that changes in one component do not ripple uncontrollably through others. New agents can be introduced by registering them within the orchestration graph and defining their task interface, without modifying the core supervisor logic. Similarly, the system supports pluggable LLMs via the LLM Gateway, allowing easy replacement or addition of model providers, whether cloud-based or on-premise, through configuration changes alone. Policy updates, including RBAC rules, sandbox constraints, or caching strategies, are isolated to the supporting infrastructure layer and can be revised independently of execution logic. This modularity not only accelerates experimentation and integration but also ensures long-term maintainability and scalability of the system in diverse deployment environments.

%% file: text/code_gene.tex
\subsubsection{Code Generator Agent: Workflow Architecture}
\begin{figure}[htbp]
\centering
\includegraphics[width=0.25\linewidth]{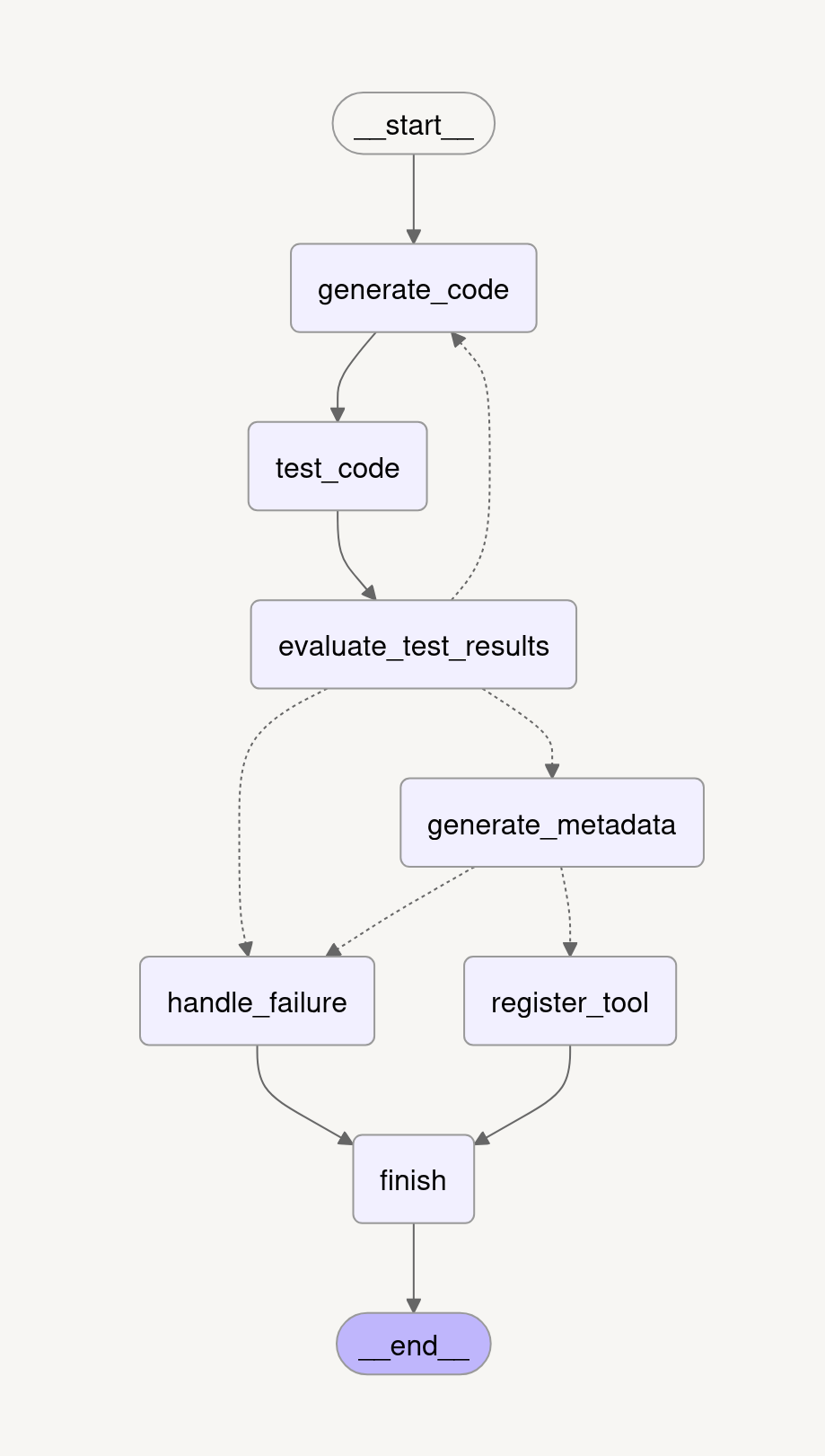}
\caption{Architecture of the Code Generator Agent. The workflow begins by generating Python code in response to a natural language request. The generated script is then sandbox-tested, and the output is evaluated for correctness, structure, and security. If validation succeeds, metadata is extracted and the tool is registered into the system for future use. Any failure encountered during the process is routed to a dedicated handler for diagnostics and recovery. This architecture enables secure, auditable, and adaptive synthesis of Kubernetes tools on demand.}
\label{fig:codegene}
\end{figure}
The \textbf{Code Generator Agent}, located within the Agent and Tool Execution Layer, enables the system to dynamically synthesize new Kubernetes tools at runtime. This agent is invoked when a user query requires functionality that does not yet exist in the predefined toolset. To ensure reliability, security, and operational correctness, it follows a structured multi-stage pipeline modeled as a LangGraph-based finite state machine (Fig.~\ref{fig:codegene}). The workflow consists of six primary components: \texttt{generate\_code}, \texttt{test\_code}, \texttt{evaluate\_test\_results}, \texttt{generate\_metadata}, \texttt{register\_tool}, and \texttt{handle\_failure}.

\paragraph{\texttt{generate\_code}}

This component generates candidate Python code based on a natural language task description. A prompt-engineered LLM is instructed to produce a complete script containing a function definition with scoped imports, robust exception handling, and a JSON-serializable output structure. Inputs to this phase include the user query, cluster context, and prior agent memory.

Using a standalone generation phase facilitates reuse of prompt templates, simplifies debugging, and allows integration of context-aware retrieval modules to further refine generation quality.

\paragraph{\texttt{test\_code}}

The generated script is then executed in a sandboxed environment (e.g., Python REPL) to validate its runtime behavior. The system checks for syntax validity, unresolved imports, runtime exceptions, and adherence to security constraints, such as restricted use of system-level APIs.

Sandboxed execution isolates unsafe code and prevents unintentional side effects. It also ensures that failure states are safely contained and reproducible.

\paragraph{\texttt{evaluate\_test\_results}}

This component is responsible for interpreting the output of the execution phase to determine whether the generated code satisfies both functional and structural requirements. The evaluation process includes verifying that the output is a well-formed JSON object, adheres to the expected response schema (e.g., \texttt{\{"status": "success", "data": \ldots\}}), and semantically aligns with the original user request. It also checks for the presence of required syntactic markers—such as function delimiters—and compliance with formatting conventions expected by the downstream system.

In addition, the evaluation phase inspects the output for unintended side effects or prohibited behaviors, such as unauthorized file access, unbounded resource consumption, or interaction with sensitive APIs. By decoupling evaluation from execution, the architecture enables modular policy enforcement and extensibility. Additional validation techniques—such as symbolic execution or mutation testing—can be incorporated independently of the execution engine. This separation enhances system robustness and ensures that only secure, interpretable, and operationally compliant tools proceed to registration.




\paragraph{\texttt{generate\_metadata}}

Upon successful validation, this component extracts structured metadata, including the function name, input schema (Pydantic class), tool variable name, and a human-readable tool description. These artifacts are essential for registration and discovery.

By generating well-formed metadata, the system ensures interoperability with LangChain tools and guarantees that each generated tool can be audited, versioned, and invoked programmatically.

\paragraph{\texttt{register\_tool}}

The validated tool is appended to a persistent Python module and registered in the global tool list using a controlled insertion strategy. This makes the new tool immediately available to other agents and future workflows.

This integration ensures system extensibility while enforcing idempotency and traceability. Each generated tool is auditable and marked as LLM-produced, supporting governance and rollback if needed.

\paragraph{\texttt{handle\_failure}}

If any of the preceding steps fail, this component handles graceful degradation. It logs the failure context, preserves the LLM inputs/outputs for future analysis, and may initiate a retry or fallback procedure.

Separating failure-handling logic ensures system robustness and enables post-hoc diagnostics, LLM fine-tuning, or human-in-the-loop oversight.

\paragraph{\texttt{finish}}
A final node that consolidates both successful (\texttt{register\_tool}) and unsuccessful (\texttt{handle\_failure}) paths, ensuring that the workflow terminates in a consistent state with proper cleanup and checkpoint persistence.

\paragraph{Separation of Execution and Evaluation}
The design choice to separate \texttt{test\_code} (execution) from \texttt{evaluate\_test\_results} (analysis) reflects best practices in secure and explainable AI engineering. It enables security through isolation, interpretability via structured evaluation logic, and adaptability for various evaluation strategies. This architectural decision also facilitates future integration of more advanced validation techniques such as mutation testing or type checking.

Overall, the Code Generator Agent ensures that LLM-generated tools are syntactically correct, semantically valid, secure, and operationally compliant. This aligns with KubeIntellect’s broader goal of safe and intelligent automation in Kubernetes environments.

%% file: text/implementation.tex
\section{Implementation}
\label{sec:implementation}

The system integrates multiple advanced components to enable intelligent, secure, and scalable Kubernetes management. The implementation follows a modular architecture composed of distinct orchestration, execution, and infrastructure layers.

\subsection{LLM-Orchestrated Technologies}

The following technologies form the backbone of intelligent task execution and dynamic orchestration:

\textbf{LangChain} is used for prompt engineering and model abstraction. The \texttt{ChatOpenAI} module integrates the GPT-4o model to translate natural language queries into structured instructions and responses.

\textbf{LangGraph} provides a graph-based finite state machine (FSM) for defining and executing workflows. Each node represents a task (e.g., log retrieval or code execution), while transitions are handled by a supervisor node based on task outcomes.

\textbf{OpenAI GPT-4o / Azure OpenAI} models serve as the core natural language processing engine. They are responsible for interpreting user queries, generating task plans, and enabling adaptive decision-making during workflow execution.

\textbf{Kubernetes Python Client} enables comprehensive interaction with the Kubernetes API, supporting resource retrieval, creation, modification, and deletion. It also facilitates advanced operations such as executing commands within pods, streaming logs and events, managing custom resources, and handling real-time updates. This client acts as the primary bridge for observing and controlling cluster state.

\textbf{PythonREPL} supports dynamic code execution by enabling agents to evaluate user-defined Python scripts in a controlled runtime, enhancing extensibility.

Together, these components form a cohesive system where natural language inputs are processed, interpreted, and mapped to executable Kubernetes operations via a hybrid graph–agent architecture.

\subsection{System Modules and Logic}

Beyond the core technologies, several system modules orchestrate intelligent behavior:

\begin{itemize}
    \item \textbf{API Interface:} The system exposes an OpenAI-compatible REST API via FastAPI. A versioned router is implemented, including endpoints such as \texttt{/chat/completions} and \texttt{/health} for workflow invocation and readiness checks.
    
    \item \textbf{Configuration Management:} Environment settings and runtime parameters (e.g., LLM provider, SSH tunnel, memory scope) are defined via Pydantic models and loaded at startup.

    \item \textbf{Kubernetes Service Layer:} A service abstraction layer encapsulates cluster interactions, including retries, logging, and SSH-based secure access. It supports both in-cluster and external configuration contexts.

    \item \textbf{Agent and Tool Definitions:} To enable modular and interpretable orchestration, the system employs a set of specialized agents, each responsible for a distinct category of Kubernetes operations. These agents encapsulate reusable toolsets aligned with functional areas such as logging, configuration, metrics, security, and lifecycle management. The underlying tools are implemented as LangChain \texttt{StructuredTool} objects, which abstract Kubernetes API calls and are dynamically reloadable at runtime. Table~\ref{tab:agent_list} summarizes the agents, their operational scope, and example tools they expose to the orchestrator for task execution.

    \item \textbf{Workflow Orchestration:} Task flows are defined in a LangGraph-based state machine. The LLM Supervisor orchestrates execution, routing requests among agents such as \texttt{DirectKubernetesTask}, \texttt{Logs}, \texttt{Security}, or \texttt{CodeGenerator}.

    \item \textbf{Checkpointing:} Human-in-the-loop (HITL) workflows are supported using a Postgres-backed checkpointing mechanism. This allows for interruption, review, and continuation of code generation steps.

    \item \textbf{LibreChat Interface:} The system integrates with LibreChat to provide a user-friendly web-based chat interface. User queries are submitted through a conversational UI and forwarded to the FastAPI backend using an OpenAI-compatible schema, enabling seamless interaction with the LLM-powered orchestration pipeline.

\end{itemize}

This implementation ensures modularity, extensibility, and traceability across the system. The use of LLMs for intelligent control flow, coupled with real-time Kubernetes access and HITL checkpoints, enables a robust and explainable management framework.


\begin{table}[]
\centering
\caption{List of specialized agents in KubeIntellect, each responsible for a distinct category of Kubernetes operations. The agents encapsulate modular toolsets aligned with Kubernetes API verb groups and operational domains, enabling fine-grained, explainable orchestration.}
\resizebox{\textwidth}{!}{%
\begin{tabular}{lll}
\hline
\textbf{Agent Name}     & \textbf{Description}                                                                                                        & \textbf{Example Tools}                                                                                         \\ \hline
\textbf{Logs}     & \begin{tabular}[c]{@{}l@{}}Handles retrieval and filtering of logs and\\ events for pods, nodes, and system.\end{tabular}   & \begin{tabular}[c]{@{}l@{}}get\_pod\_logs, watch\_events, \\ list\_namespace\_events\end{tabular}              \\
\textbf{Configs}  & \begin{tabular}[c]{@{}l@{}}Inspects and manages declarative \\ configurations of core workloads and services.\end{tabular}  & \begin{tabular}[c]{@{}l@{}}list\_deployments, get\_service\_config, \\ validate\_configmap\end{tabular}        \\
\textbf{RBAC}           & \begin{tabular}[c]{@{}l@{}}Audits access control policies\\ and role bindings.\end{tabular}                                 & \begin{tabular}[c]{@{}l@{}}list\_roles, check\_role\_binding, \\ inspect\_service\_account\end{tabular}        \\
\textbf{Metrics}        & \begin{tabular}[c]{@{}l@{}}Gathers resource and application\\ performance metrics.\end{tabular}                             & \begin{tabular}[c]{@{}l@{}}get\_cpu\_usage, get\_node\_metrics,\\ pod\_network\_io\end{tabular}                \\
\textbf{Security}       & \begin{tabular}[c]{@{}l@{}}Analyzes cluster security posture and\\ detects violations.\end{tabular}                         & \begin{tabular}[c]{@{}l@{}}analyze\_audit\_logs, check\_psp\_violations,\\ scan\_vulnerabilities\end{tabular}  \\
\textbf{Lifecycle}      & \begin{tabular}[c]{@{}l@{}}Executes control actions on\\ workloads and nodes.\end{tabular}                                  & \begin{tabular}[c]{@{}l@{}}scale\_deployment, cordon\_node,\\ restart\_pod, evict\_pod\end{tabular}            \\
\textbf{Execution}      & \begin{tabular}[c]{@{}l@{}}Runs interactive commands inside\\ containers and tunnels.\end{tabular}                          & exec\_in\_pod,port\_forward, attach\_container                                                                 \\
\textbf{Deletion}       & \begin{tabular}[c]{@{}l@{}}Deletes individual or bulk\\ Kubernetes resources.\end{tabular}                                  & \begin{tabular}[c]{@{}l@{}}delete\_pod, cleanup\_jobs,\\ delete\_namespace\_resources\end{tabular}             \\
\textbf{AdvancedOps}    & \begin{tabular}[c]{@{}l@{}}Handles complex operations like binding, \\ server-side apply, or CRD interactions.\end{tabular} & \begin{tabular}[c]{@{}l@{}}apply\_manifest, bind\_pod, approve\_certificate,\\ finalize\_resource\end{tabular} \\
\textbf{Code Generator} & \begin{tabular}[c]{@{}l@{}}Synthesizes and registers new tools\\ dynamically when no predefined tool matches.\end{tabular}  & \begin{tabular}[c]{@{}l@{}}generate\_k8s\_script, register\_dynamic\_tool,\\ test\_code\_snippet\end{tabular}  \\ \hline
\end{tabular}
\label{tab:agent_list}
}
\end{table}

%% file: text/evaluation.tex
\section{Evaluation}
\label{sec:evaluation}

To assess the performance, responsiveness, and scalability of the KubeIntellect system under realistic operational conditions, we conducted evaluation experiments in a live Kubernetes cluster consisting of four nodes: a dedicated control plane and three worker nodes (See Table \ref{tab:testbed}). Each node was provisioned with 32 vCPUs and 500 GB of storage, while RAM configurations included 64 GB on three nodes and 128 GB on the largest worker. The cluster hosted 170 pods distributed across 18 namespaces, with additional components including 71 deployments, 49 stateful sets, 186 services, and over 100 ConfigMaps and Secrets—representing a diverse and production-grade workload. This environment provided a representative testbed to validate the system’s ability to process user queries, orchestrate workflows, and maintain consistent behavior under complex, multi-agent workloads. The evaluation focused on measuring latency, resource usage, and orchestration efficiency across the system’s modular components and LLM-driven logic.

\subsection{Latency Analysis Across System Layers and Code Generation Workflow}
\input{text/table_latency}
To evaluate the runtime behavior and responsiveness of the KubeIntellect system, we measured the latency of core components across two dimensions: (1) high-level architectural layers—\texttt{Supervisor}, \texttt{Agent}, and \texttt{Tool}, and (2) the internal stages of the Code Generator Agent. The latency statistics are summarized in Table~\ref{tab:latency_stats}, which reports the number of invocations, average latency, and standard deviation for each component.

At the architectural level, the \textbf{Supervisor} tier—responsible for task orchestration and prompt routing—was the most frequently invoked (549 calls), with a mean latency of 0.445 seconds (SD = 1.034). This suggests that the LLM-based control logic introduces minimal decision-making overhead. The \textbf{Agent} tier, encompassing Configs, Metrics, and Security subsystems (excluding the Code Generator), showed higher variability with a mean latency of 2.625 seconds (SD = 3.059), driven by Kubernetes API access and inter-agent coordination. The \textbf{Tool} tier, composed of atomic operations such as \texttt{get\_pod\_logs}, \texttt{list\_namespaces}, and \texttt{check\_loadbalancer\_external\_ips}, demonstrated the lowest average latency at 0.111 seconds (SD = 0.263), confirming their suitability for rapid and reusable execution.

In parallel, the bottom half of Table~\ref{tab:latency_stats} provides a fine-grained breakdown of the \textbf{Code Generator Agent} workflow. As expected, the \texttt{generate\_code} stage was the most time-consuming, with a mean latency of 8.416 seconds (SD = 5.323), largely attributable to LLM-based code synthesis. In contrast, downstream stages such as \texttt{test\_code} (0.088 s), \texttt{evaluate\_test\_results} (0.003 s), and \texttt{register\_tool} (0.011 s) incurred negligible cost, thanks to the use of in-process REPLs and file-based persistence. Metadata generation, another LLM-dependent step, averaged 2.625 seconds (SD = 1.145), while failure handling and termination completed almost instantaneously (0.001 s).

Overall, these results validate the efficiency of the system’s modular design. While LLM inference dominates latency within the Code Generator Agent, other components across the architecture demonstrate fast, predictable, and interactive performance—making KubeIntellect suitable for real-time Kubernetes management at scale.

\subsection{Query-Level Evaluation and Behavior Analysis}

To assess the end-to-end orchestration behavior of the KubeIntellect system, we executed a total of \textbf{200 diverse natural language queries}, each representing a distinct user intent across various operational domains. Among these, 77 queries did not match any existing registered tool and therefore triggered the Code Generator Agent to synthesize new tools dynamically. Of these, 63 tools were successfully generated, validated, and registered into the system. The remaining 14 tool generation attempts failed after three retries. These failures were later resolved through improvements in prompt engineering and the tool registration process.

In addition to tool synthesis (which require multi-agent task execution), approximately 30 other queries required complex, multi-agent workflows. For instance, certain queries first invoked the Configs agent to retrieve a list of namespaces, and subsequently triggered the Logs or Metrics agent to fetch resources per namespace. These scenarios demonstrate the Supervisor’s ability to dynamically sequence and compose agent calls across different domains in response to a single user query.

Aside from the 14 tool generation failures, all other queries returned valid, context-aware responses without hallucinations or misinterpretations. This corresponds to a \textbf{93\% success rate} in tool creation and a \textbf{100\% success rate} for queries executed via existing agents and tools.

The system also employed human-in-the-loop (HITL) clarification to enhance reliability and safety during ambiguous or under-specified interactions. Of the total queries, 40 required initial clarification from the user at the point of submission. Additionally, 118 agent responses triggered clarification prompts or human approval checkpoints during intermediate orchestration stages, including those related to tool creation, registration, or iterative validation.

In 10 cases, the Supervisor explicitly rejected the result returned by an agent and either retried the task or forwarded it to an alternate agent. This behavior highlights the system’s adaptive control logic and built-in recovery capabilities.

Earlier versions of the system encountered two key limitations: hallucinations in language model responses and improper reuse of stale memory (such as outdated metrics). These issues were resolved in the current release through targeted improvements to prompt construction, orchestration rules, and memory invalidation logic. No such issues were observed during the evaluation experiments reported in this study.

\subsection{Computational Resource Footprint Under Load}
To evaluate the runtime efficiency and scalability of KubeIntellect in multi-user scenarios, we measured the CPU and memory consumption of its core components—the backend orchestration system and the browser-based frontend—under three workload conditions: \textit{Idle} (no active users), \textit{1 User}, and \textit{2 Users}. Resource usage metrics were collected using Kubernetes-native instrumentation across both the \texttt{mksa} backend namespace and the frontend deployment.

Table~\ref{tab:resource_usage} presents the observed resource footprint in terms of CPU cores and memory usage (MiB). In the idle state, the backend consumed only \textbf{0.01 cores} and \textbf{240 MiB} of memory, demonstrating minimal standby overhead. The frontend, based on a React application with WebSocket support, maintained a baseline of \textbf{0.03 cores} and \textbf{487 MiB}, reflecting static asset caching and persistent UI rendering.

As user interactions increased, the backend’s CPU usage scaled linearly, rising to \textbf{0.15 cores} for one active user and \textbf{0.30 cores} for two users. This growth corresponds to increased agent activation, prompt evaluation, and tool orchestration. In contrast, backend memory usage remained stable across load levels, fluctuating only slightly from \textbf{240 MiB} to \textbf{263 MiB}, indicating efficient memory management during concurrent execution.

Frontend CPU usage also grew with user load, increasing from \textbf{0.03} to \textbf{0.08} and \textbf{0.10 cores}, primarily due to user interface updates and message streaming over WebSockets. Memory usage remained bounded, with only minor increases from \textbf{487 MiB} to \textbf{510 MiB}, confirming the frontend's predictable memory profile even under concurrent user sessions.

Overall, these results demonstrate that KubeIntellect maintains a low and predictable computational footprint across usage conditions. Its modest idle resource requirements and graceful scaling under multiple users make it suitable for deployment in constrained environments such as developer workstations, academic cloud infrastructure, and edge clusters.


\begin{table}[ht]
\centering
\caption{Resource usage of KubeIntellect backend and frontend under varying user load. CPU is reported in virtual cores (vCPU), and memory in mebibytes (MiB).}
\label{tab:resource_usage}
\begin{tabular}{lcccc}
\toprule
\multirow{2}{*}{\textbf{User Load}} & \multicolumn{2}{c}{\textbf{Backend}} & \multicolumn{2}{c}{\textbf{Frontend}} \\
\cmidrule(lr){2-3} \cmidrule(lr){4-5}
 & \textbf{CPU [vcores]} & \textbf{Memory [MiB]} & \textbf{CPU [vcores]} & \textbf{Memory [MiB]} \\
\midrule
Idle       & 0.01 & 240 & 0.03 & 487 \\
1 User     & 0.15 & 260 & 0.08 & 503 \\
2 Users    & 0.30 & 263 & 0.10 & 510 \\
\bottomrule
\end{tabular}
\end{table}


\begin{table}[ht]
\centering
\caption{Hardware Configuration and Kubernetes Cluster Component Summary}\caption{%
Hardware and Kubernetes Cluster Configuration. The top half of the table summarizes the hardware resources provisioned for the evaluation cluster. The bottom half provides a breakdown of the cluster's components at the time of testing, including the number of namespaces, pods, deployments, stateful sets, services, and associated Kubernetes resources. 
}
\label{tab:testbed}
\renewcommand{\arraystretch}{1.2}
\begin{tabular}{lccc}
\toprule
\multicolumn{4}{c}{\textbf{Hardware Resources}} \\
\midrule
\textbf{Instance Name} & \textbf{vCPUs} & \textbf{Disk} & \textbf{RAM} \\
\midrule
ControlPlane & 32 & 500 GB & 64 GB \\
Worker\_1     & 32 & 500 GB & 64 GB \\
Worker\_2     & 32 & 500 GB & 64 GB \\
Worker\_3     & 32 & 500 GB & 128 GB \\
\midrule
\multicolumn{4}{c}{\textbf{Kubernetes Cluster Component Summary}} \\
\midrule
\textbf{Component} & \textbf{Count} & & \\
\midrule
Nodes            & 4   & & \\
Namespaces       & 18  & & \\
Pods             & 170 & & \\
Deployments      & 71  & & \\
StatefulSets     & 49  & & \\
DaemonSets       & 5   & & \\
Services         & 186 & & \\
PersistentVolumes (PV)         & 55  & & \\
PersistentVolumeClaims (PVC)   & 55  & & \\
ConfigMaps       & 121 & & \\
Secrets          & 124 & & \\
ServiceAccounts  & 147 & & \\
\bottomrule
\end{tabular}
\end{table}

\subsection{Challenges and Observations}
The proposed system demonstrates robust capabilities in simplifying Kubernetes management; however, certain limitations and challenges require further investigation. One significant limitation is the reliance on OpenAI's cloud-based LLMs, which raises concerns about data security and privacy. For Kubernetes clusters managing sensitive or confidential data, transmitting queries and responses to external servers can introduce compliance risks. This underscores the need for adopting on-premises or local LLM models that ensure all computations and data handling occur within controlled environments. Additionally, utilizing local LLMs can mitigate the operational costs associated with cloud-based API usage, making the system more accessible for organizations with high query volumes or continuous monitoring requirements. Since the framework is developed using LangGraph, its compatibility with local LLM deployments provides a feasible path for this transition. However, further evaluation of local LLM models is necessary to validate their performance, scalability, and integration within the existing architecture.

Another critical challenge lies in the use of the Python REPL executor, which provides dynamic flexibility for executing Python code and handling unanticipated tasks. While this feature enhances adaptability, it introduces potential security vulnerabilities, such as arbitrary code execution, data leakage, denial of service (DoS) attacks through resource exhaustion, and the persistence of malicious code. Mitigating these risks requires a comprehensive approach, including sandboxing the execution environment, enforcing strict access controls via role-based access control (RBAC), implementing input validation, and restricting system-level access. Furthermore, integrating predefined commands and templates in place of free-form code execution can significantly reduce the attack surface, ensuring a balance between flexibility and security.

During the system evaluation, several observations related to task execution and dependency management were made. Complex multi-agent workflows occasionally defaulted to the Code Generator Agent for unanticipated tasks, resulting in increased execution times. Future evaluations should compare the computational cost and reliability of using pre-built tools versus dynamically generated code to address such tasks. Additionally, metrics retrieval tasks were found to depend on external configurations, such as the availability of the Kubernetes Metrics Server. This highlights the importance of proper cluster setup to ensure seamless operation and uninterrupted functionality.

Despite these challenges, the system’s ability to interpret human queries, dynamically orchestrate tasks, and interact with the Kubernetes API was validated, confirming its potential as a robust and adaptive solution for Kubernetes management. Further investigations will focus on optimizing task execution strategies, expanding scalability testing, and addressing security and privacy concerns to enhance the system’s utility and resilience.

%% file: text/table_latency.tex
\begin{table}[]
\centering
\caption{Latency statistics for core execution components in KubeIntellect. 
The top subtable summarizes latency metrics across system layers (\texttt{Supervisor}, \texttt{Agent}, and \texttt{Tools}), 
while the bottom subtable presents detailed latency measurements for each stage of the Code Generator Agent workflow.}
\begin{tabular}{lccc}
\hline
\multicolumn{4}{l}{\textbf{KubeIntellect Stages Latency Statistics}}        \\ \hline
run\_name                       & count        & mean         & std         \\ \hline
Supervisor                      & 738          & 0.445        & 1.034       \\
Agent                           & 343          & 2.625        & 3.059       \\
Tools                           & 226          & 0.111        & 0.263       \\ \hline
\multicolumn{4}{l}{\textbf{Code Generator Stages Latency Statistics}} \\ \hline
run\_name                       & count        & mean         & std         \\ \hline
generate\_code                  & 117          & 8.416        & 5.323       \\
test\_code                      & 117          & 0.088        & 0.270       \\
evaluate\_test\_results         & 117          & 0.003        & 0.001       \\
generate\_metadata              & 63           & 2.625        & 1.145       \\
register\_tool                  & 63           & 0.011        & 0.004       \\
handle\_failure                 & 14           & 0.001        & 0.001       \\
finish                          & 77           & 0.001        & 0.001       \\ \hline
\end{tabular}
\label{tab:latency_stats}
\end{table}

%% file: text/demo.tex
\section{Code Availability and Demo Infrastructure}
\label{sec:availability}
The source code for KubeIntellect is publicly available under an open repository at:

\begin{center}
\url{https://github.com/MSKazemi/KubeIntellect}
\end{center}

This repository includes all core components of the platform, including the LLM-powered orchestration engine, agent definitions, tool interfaces, and the FastAPI-based REST interface. It also provides detailed instructions for setup, development, and extension.

\vspace{0.5em}
\subsection*{Azure-Based Demo Environment}

To facilitate testing and experimentation, a separate infrastructure repository provides a fully automated deployment pipeline for launching a demo environment on Microsoft Azure. This testbed uses \textbf{Terraform} to provision an \textbf{Azure Kubernetes Service (AKS)} cluster, configure monitoring dashboards, and deploy KubeIntellect components without requiring manual intervention.

The deployment workflow is designed for reproducibility and minimal user effort. The only prerequisites are:
\begin{itemize}
    \item An active Azure subscription
    \item The Azure CLI
    \item Terraform (v1.3+)
    \item A Bash shell and \texttt{curl}
\end{itemize}

After cloning the infrastructure repository, users can run:

\begin{verbatim}
./validate-environment.sh
./deploy-automated.sh
\end{verbatim}

The first script checks environment prerequisites and gathers Azure credentials. The second deploys the full KubeIntellect stack, including:
\begin{itemize}
    \item AKS cluster provisioning with autoscaling
    \item NGINX ingress controller setup
    \item Prometheus and Grafana dashboards via \texttt{kube-prometheus-stack}
    \item LLM backend configuration and endpoint exposure
    \item Automatic \texttt{kubectl} context and ingress configuration
\end{itemize}

At the end of deployment, access credentials and URLs for the API gateway and dashboards are printed, enabling immediate interaction with the system. This infrastructure is ideal for reproducible benchmarking, demos, and developer onboarding.

\vspace{0.5em}
\subsection*{Support for Local Testing via \texttt{kind}}

To broaden accessibility beyond cloud environments, work is ongoing to support a lightweight local testbed using \textbf{\texttt{kind}} (Kubernetes in Docker). This will enable rapid evaluation of KubeIntellect on developer machines without cloud infrastructure, while maintaining compatibility with the same orchestration and agent logic. Instructions for launching the local demo environment will be added to the repository in an upcoming release.

%% file: text/appendix.tex
\appendix
\section{Monitoring Data Types in Kubernetes}
\label{appendix:monitoring_data}

\input{text/t1}
Table \ref{tab:k8s_monitoring} provides a comprehensive summary of the key data types used for monitoring and managing Kubernetes clusters. The data is categorized into four distinct groups: \textit{Logs (Operational Data)}, \textit{Configurations (Static Data)}, \textit{Metrics (Quantitative Data)}, and \textit{Security (Audit and Compliance Data)}. Each category highlights its focus, key items, and primary purpose within Kubernetes monitoring workflows.

The \textit{Logs (Operational Data)} category captures real-time system activity and runtime issues. This includes system logs (e.g., API server, Scheduler), pod logs for application behavior, and node logs for resource conditions. Key events such as scheduling failures, warnings, and errors like \verb|ImagePullBackOff| or \verb|CrashLoopBackOff| are monitored to provide insights for debugging and tracking system behavior.

The \textit{Configurations (Static Data)} category contains declarative information about the desired state of the cluster. This includes pod specifications, service configurations, and networking setups like ingress rules and network policies. Other critical items, such as Secrets, ConfigMaps, RBAC roles, and persistent storage configurations, are essential for ensuring system compliance, troubleshooting, and auditing.

The \textit{Metrics (Quantitative Data)} category focuses on tracking resource usage and performance trends. Metrics such as CPU, memory, disk usage, and network traffic provide the basis for capacity planning and optimization. Storage metrics, application-specific KPIs, and job execution metrics are also monitored to identify underutilization or bottlenecks.

Finally, the \textit{Security (Audit and Compliance Data)} category monitors security events and access control changes to protect the cluster from potential breaches. It includes audit logs, RBAC changes, network policy violations, and vulnerability scans. This data ensures compliance with security policies and supports proactive threat detection.

This table serves as a structured guide for understanding the diverse types of monitoring data in Kubernetes, emphasizing their roles in maintaining cluster health, performance, and security.

%% file: text/t1.tex
\begin{table*}[htbp]
\caption{Summary of Monitoring Data Types in Kubernetes}
\centering
\resizebox{\textwidth}{!}{%
\begin{tabular}{|p{3cm}|p{4cm}|p{6cm}|p{4cm}|}
\hline
\textbf{Type of Data} & \textbf{Focus} & \textbf{Key Items} & \textbf{Purpose} \\ \hline

\textbf{Logs (Operational Data)} & 
Captures real-time system activity and runtime issues. & 
\begin{itemize}
    \item System Logs: API server, Scheduler, Controller Manager, etcd logs.
    \item Pod Logs: Application logs, error/crash logs.
    \item Node Logs: Disk pressure, memory issues.
    \item Events: Scheduling failures, pod lifecycle events.
    \item Warnings and Errors: Resource issues, \texttt{ImagePullBackOff}, \texttt{CrashLoopBackOff}.
\end{itemize} & 
Provides insights for debugging, real-time issue detection, and tracking system behavior. \\ \hline

\textbf{Configurations (Static Data)} & 
Contains declarative information about the cluster’s desired state. & 
\begin{itemize}
    \item Pod Specifications: Resource limits, volumes.
    \item Service Configurations: LoadBalancer, ClusterIP, NodePort.
    \item Networking: Ingress rules, network policies.
    \item Deployments: Replica counts, update strategies.
    \item Secrets/ConfigMaps: Sensitive data and environment configurations.
    \item RBAC: Roles, bindings.
    \item Persistent Storage: PVs, PVCs.
    \item CronJobs/Jobs: Scheduling, resource allocation.
\end{itemize} & 
Ensures system compliance, auditing, and troubleshooting configuration issues. \\ \hline

\textbf{Metrics (Quantitative Data)} & 
Tracks resource usage and system performance trends. & 
\begin{itemize}
    \item Resource Metrics: CPU, memory, disk usage.
    \item Node and Pod Metrics: Resource requests and limits.
    \item Network Metrics: Traffic, latency.
    \item Storage Metrics: PV usage, IOPS.
    \item Application Metrics: Custom KPIs.
    \item Job Metrics: Success/failure rates, execution times.
\end{itemize} & 
Enables capacity planning, performance optimization, and identifying over/underutilization. \\ \hline

\textbf{Security (Audit and Compliance Data)} & 
Monitors security events and access control changes. & 
\begin{itemize}
    \item Audit Logs: API server actions, access logs.
    \item RBAC Logs: Role/RoleBinding changes.
    \item Pod Security Policies (PSP): Violations, enforcement.
    \item Network Policies: Logs for unauthorized access attempts.
    \item Vulnerability Scans: Container, pod, and node assessments.
\end{itemize} & 
Protects the cluster from security breaches and ensures compliance with security policies. \\ \hline

\end{tabular}
\label{tab:k8s_monitoring}
}
\end{table*}